\documentclass[preprint,12pt]{elsarticle}

\usepackage{graphicx}
\usepackage{amssymb}

\usepackage{float}

\journal{Journal of Artificial Intelligence and Soft Computing Research
}

\begin{document}

\begin{frontmatter}


\title{A study for Image compression using Re-Pair algorithm}

\author[1]{Pasquale De Luca\corref{correspondingauthor}}
\ead{deluca@ieee.org}

\author[2]{Vincenzo Maria Russiello}
\author[2]{Raffaele Ciro Sannino}
\author[2]{Lorenzo Valente}

\cortext[correspondingauthor]{Corresponding Author}
\address[1]{University of Naples "Parthenope", Department of Science and Technologies, Centro Direzionale C4, Napoli I-80143}
\address[2]{University of Salerno, Department of Computer Science, via Giovanni Paolo II, Fisciano I-84084}
\begin{abstract}
The compression is an important topic in computer science which allows we to storage more amount of data on our data storage. There are several techniques to compress any file. In this manuscript will be described the most important algorithm to compress images such as JPEG and it will be compared with another method to retrieve good reason to not use this method on images. So to compress the text the most encoding technique known is the Huffman Encoding which it will be explained in exhaustive way. In this manuscript will showed how to compute a text compression method on images in particular the method and the reason to choice a determinate image format against the other. The method studied and analyzed in this manuscript is the Re-Pair algorithm which is purely for grammatical context to be compress. At the and it will be showed the good result of this application.
\end{abstract}


\end{frontmatter}


\section{Introduction}
\label{S:1}
This manuscript is a deep study so it is possible execute and use a particular compression algorithm on images. Aside the canonical compression algorithm for images such as JPEG\citep{prejpeg} and PNG, this work has the aim to show how to apply the \textbf{Re-Pair} algorithm, purely for text compression, on images. We have analyzed in literature if is present a similar or related work but the search has worst results so there are mainly paper and works on Re-Pair applied on the text. We would show how is possible execute the latter algorithm on images using methods and techniques to generate a good input for the Re-Pair algorithm which is purely for text segmentation justified and structured according to canonical grammar rules.
It is done a brief analysis of the problem about the compression of images using several and canonical algorithm indeed after we will describe the method used to compute in efficient way this technique so that for future works can be used, in particular great importance it will be given to compression method for images to compare the different techniques to reveal which method is the better so that it will be a correct choice to have a very efficient result computed by operation of the \textit{Re-Pair} algorithm.

\section{Methods}
\label{sec:methods}
In this section we will show the method can compute \textbf{Re-Pair} algorithm on images, in particular the techniques used to convert an any image from any graphical format in \textbf{BitMap} format, after it will explain why it used this format.
We can classify compression algorithm in two groups or rather \textbf{\textit{lossless and lossy}} \citep{lossy1}.

\subsection{Lossless compression methods}
Lossless compression algorithms are computing compression methods that allow the original data to be perfectly reconstructed from the compressed data, briefly it is possible compress an image without loss graphical and data information. 
Lossless algorithm such as \textit{Huffman coding} \citep{huff1}, which belongs to \textit{Entropy Encoding \textbf{subfamily}} is a most used compression method on which based a lot of compression algorithm in particular JPEG \citep{lossless1} where it is possible compress an image opening it in \textbf{binary} mode and reading a single byte like ASCII symbol and after apply Huffman Encoding to generate a compression version of raw image. Other algorithms that belong to \textit{lossless} family are:
\begin{itemize}
    \item PNG;
    \item TIFF;
    \item TGA;
    \item BPG.
\end{itemize}
Also there are other methods we cite the most importants used, indeed we found other techniques so \textbf{RLE \citep{runleght1} and Chain Codes \citep{cc1}}, aside the latter which used on monochromatic images.

\subsection{Lossy compression methods}
This kind of methods have as aim to compress an image losing the several information starting from reducing the \textit{color space} like \textbf{chroma subsampling} method and any \textit{Transform coding} which belongs an important transform such as \textbf{Discrete Cosine Transform} \citep{dct1}.

Following it is showed a figure that represent various applications of a DCT on image:

\begin{figure}[H]
\centering
\includegraphics[width=.3\textwidth,viewport=0 0 400 400]{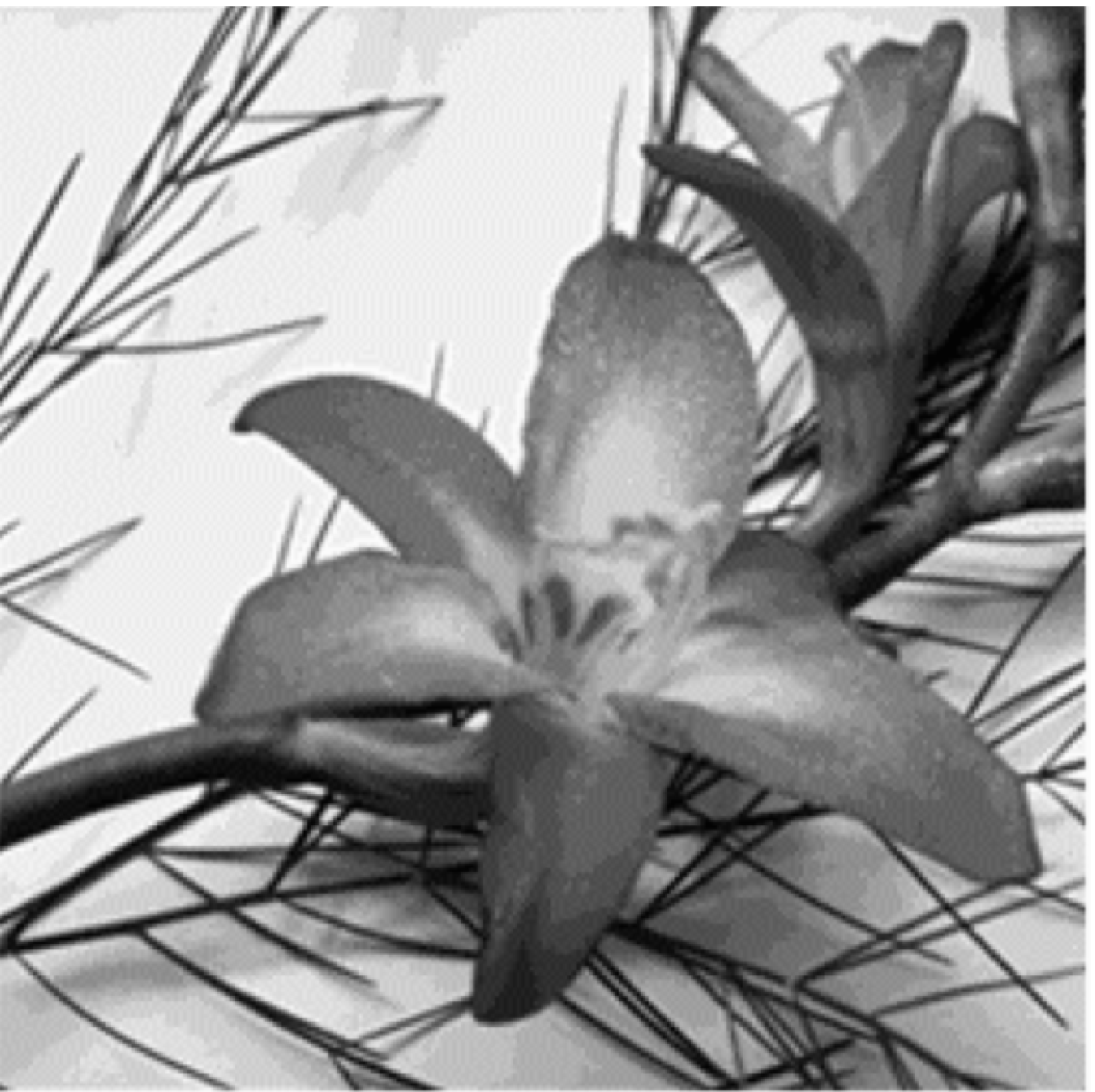}\quad
\includegraphics[width=.3\textwidth, viewport=0 0 400 400]{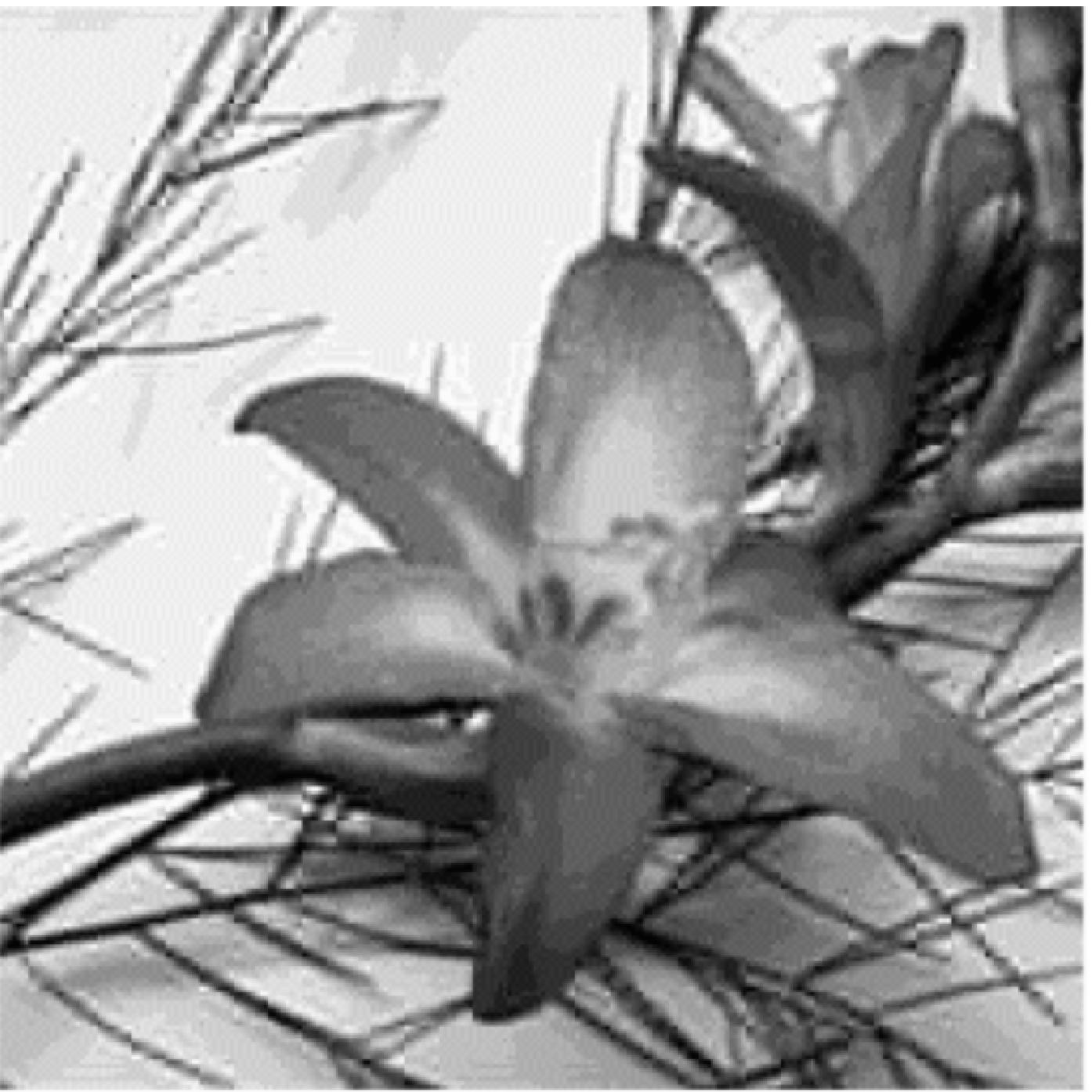}\quad
\includegraphics[width=.3\textwidth, viewport=0 0 400 400]{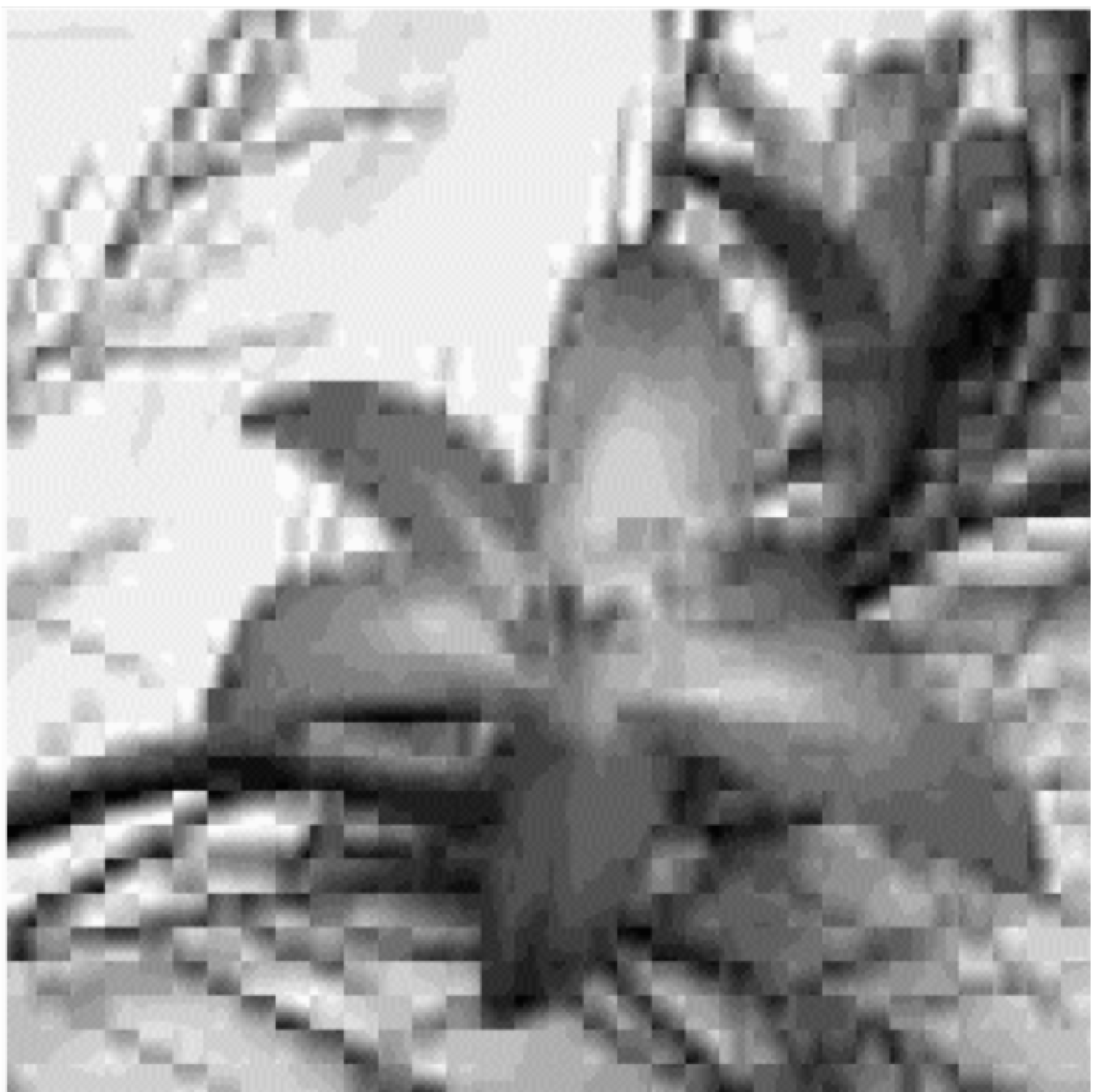}

\medskip

\caption{DCT compression step}
\label{pics:dctfig}
\end{figure}

In Figure \ref{pics:dctfig} it is possible see the differences of compressed images, the first image is the raw, the second has a 8 coefficients of compression and the last image has 64 coefficients of compression, more coefficients increase the percentage of compression and automatically increments the losing of information. Overtime are combined these two family but with minimal final result \citep{jpegdct}.
Any method explained till now has been developed in particular for images, after we introduce how to configure and use Re-Pair algorithm on images \citep{comparejpeg}.

\subsection{Re-Pair on Images}
Re-Pair is an efficient grammar compressor that operates by recursively replacing high-frequency character pairs with new grammar symbols the main approach is to replace the most frequently occurring pairs first, and for each pair add a dictionary entry mapping the new symbol to the replaced pair \citep{offline}.

Briefly we describe the step of the algorithm, the \textbf{first step} compute the \textit{sequence array} which is an array structure where each entry consists of a symbol value and two pointers which are used to create doubly linked lists between sequences of identical pairs. The \textbf{second step} consists to build a \textit{active pairs table}. This is a hash table from a pair of symbols to a pair record, which is a collection of information about that specific pair. 

The Re-Pair algorithm starts counting the pairs in input using a \textit{flag} to indicate that a pair is seen once, and if the pair is encountered again a pair record is created.

Mainly the Re-Pair algorithm works on the text in particular on correctly grammar based text for this reason is correct to say something about the \textbf{Context-Free Grammar}, now \textbf{CFG}. 

For definition:

\begin{quote}
   A context-free grammar is a collection of rules of the form: 
   \begin{center}
       $ A = x\textsubscript{1},x\textsubscript{2},x\textsubscript{3},...,x\textsubscript{k} $
   \end{center}
   where x\textsubscript{1},x\textsubscript{2},x\textsubscript{3},...,x\textsubscript{k} are either terminal symbols (letters in the alphabet) or symbols that appear on the left-hand side of some rule.
\end{quote}

To execute the Re-Pair algorithm on image we must convert any images to BMP Format why the common format such as PNG and JPEG and other already compressed format does not make it perform its compression instead a BMP or rather \textbf{Bitmap Image File} is a simple format where there is not a high level of compression indeed its size on hard disk are greater than any images of different format.
We tried to execute Re-Pair algorithm on JPEG and PNG images but the result was been very bad which are showed in following table:
\begin{table}[H]
\centering
\begin{tabular}{|r|r|r|r|}
\hline
\multicolumn{1}{|l|}{Name file} & \multicolumn{1}{l|}{Type image input} & \multicolumn{1}{l|}{Size (Kb)} & \multicolumn{1}{l|}{Size output (Kb)} \\ \hline
hello                      & JPEG                                  & 67                             & 72                                               \\ \hline
test1                      & JPEG                                  & 192                            & 388                                              \\ \hline
test1                      & PNG                                   & 86                             & 96                                               \\ \hline
test2                      & PNG                                   & 155                            & 227                                              \\ \hline
\end{tabular}
\caption{Execution of Re-Pair algorithm on different images format}
\label{tab:tab1}
\end{table}
It is clear that can not possible use already compressed images why there is confusion and redundancy into compressed images.

\subsection{Convert to BMP Format}
The main step to can apply this technique of compression is to convert any images file to BMP format so it is a simple format with low rate compression and this one allow to compression algorithm to be efficient so the redundancy in bitmap picture favors a better compression \citep{jpegbmp}.

\subsection{Convert BMP to ASCII}
Also this step is optional but to have an efficient compression with good results we recommend to covert the bitmap image to ASCII file using a simple method with any programming language, to open the image file in \textbf{binary} \citep{ascii1} mode and read chunk of octet bit, after convert the octet of bit in decimal number such as this table shows the conversion of bit to decimal to ASCII \citep{ascii2} code:

\begin{table}[H]
\centering
\begin{tabular}{|r|r|r|}
\hline
\multicolumn{1}{|l|}{Letter} & \multicolumn{1}{l|}{ASCII Code} & \multicolumn{1}{l|}{Binary} \\ \hline
a                            & 97                              & 01100001                    \\ \hline
b                            & 98                              & 01100010                    \\ \hline
c                            & 99                              & 01100011                    \\ \hline
d                            & 100                             & 01100100                    \\ \hline
\end{tabular}
\caption{ASCII Table example}
\label{tab:tab2}
\end{table}

In Table \ref{tab:tab2} is possible to see the Binary code, of the first four number of ASCII table, represented by octet of bit then transform to decimal number is possible associate the relative ASCII code using cast of value.

In \ref{sec:methods} is possible view the results after this conversions.

\section{Results}
In this section will be described briefly the result obtained after the converting of the images using the previous method. Our techniques consists to reading the BMP image using \textbf{\textit{zig-zag}} reading techniques. In this way there are more possibility of an efficient result. The following table shows the result of Re-Pair algorithm executed on BMP images:

\begin{table}[H]
\centering
\begin{tabular}{|r|r|r|}
\hline
\multicolumn{1}{|l|}{Name} & \multicolumn{1}{l|}{Size (Mb)} & \multicolumn{1}{l|}{Size out (Mb)} \\ \hline
hello                      & 4,4                            & 1,5                                \\ \hline
Ray                        & 1,1                            & 0,4                                \\ \hline
Lena                       & 2,1                            & 0,9                                \\ \hline
binary\_ex                 & 1,0                            & 0,3                                \\ \hline
\end{tabular}
\caption{Re-Pair algorithm on BMP images}
\label{tab:tab3}
\end{table}

In the table \ref{tab:tab3} it is possible to note that the rate compression is around the 70 percentage, in particular on monochromatic images why the redundancy of \textbf{chroma} pixel during the conversion to ASCII generates same characters which are the best input for the Re-Pair algorithm.

\section{Discussion}
We can note the good result in previous section indeed the it is a proposal how method to compress and hide images file using a grammar compression method. The reasoning to convert any image file to BMP and after to ASCII allows to \textit{Re-Pair} to have good final results but there is minimal problem on \textbf{linear} reading of the file why a linear reading can make problem with detecting of the pairs in text, following the structure of algorithm. A new technique to add to these step are to \textbf{split} the image in \textit{three} different image each for any \textbf{colour channel} or rather \textbf{RGB}. After to apply the Re-Pair on these file using a \textbf{Discrete Cosine Transform} reading so to have a better classification to accommodate the redundancy in the start file.

\section*{References}

\bibliographystyle{plain}

\end{document}